\documentclass[useAMS,usenatbib]{mn2e} 
\usepackage{psfig}

\def\kms{km ${\rm s}^{-1}$}

\def\ch2{$\chi^2$}
\def\dg{$^{\circ}$}

\def\kms {\hbox{${\rm km\ s}^{-1}$}}

\def\scm  {$\hbox{{\rm cm}}^{-2}$}    

\def \HI {H{\sc \,i}}

\def\lapp{\ifmmode\stackrel{<}{_{\sim}}\else$\stackrel{<}{_{\sim}}$\fi}
\def\gapp{\ifmmode\stackrel{>}{_{\sim}}\else$\stackrel{>}{_{\sim}}$\fi}

\title[Searching for molecules at $z\sim3$ with the upgraded
    ATCA]{A 22 GHz search for molecular absorption at {\boldmath $z\sim3$} with the upgraded ATCA}
    \author[S. J. Curran et
    al.]{S. J. Curran$^{1}$\thanks{E-mail: sjc@bat.phys.unsw.edu.au
    (SJC)}, M. T. Murphy$^{1}$, J. K. Webb$^{1}$ and
    Y. M. Pihlstr\"{o}m$^{2}$\\$^{1}$School of Physics, University of
    New South Wales, Sydney N.S.W. 2052,
    Australia\\$^{2}$National Radio Astronomy Observatory, Socorro, NM
    87801, USA}

\begin{document}

\date{Accepted ---. Received ---; in original form ---}

\pagerange{\pageref{firstpage}--\pageref{lastpage}} \pubyear{2002}

\maketitle

\label{firstpage}

\begin{abstract}
We report a $\lambda\gapp1$ cm search for rotational molecular
absorption towards quasars, now possible with the upgraded Australia
Telescope Compact Array (ATCA). The targets were PKS 0201+113, PKS
0336--017 and Q 0537--286, where known damped Lyman-alpha absorption
systems (DLAs) could cause redshifted molecular absorption in the 12
mm band of the telescope. We place $3\sigma$ upper limits on any
HCO$^+$ $0\rightarrow1$ absorption features of $<30$ mJy per 3.4 \kms
~channel.  The non-detections could be attributed to the inherent low
metallicities in DLAs leading to generally low H$_2$, and thus
HCO$^+$, column densities. In general, the detection of molecular
rotational transitions in DLAs could be further hindered by a
lower than expected CO-to-H$_2$ conversion ratio, whether due either to 
photoionization of carbon or its relative under-abundance at high
redshift. 
\end{abstract}


\section{Introduction}\label{sec:intro}

Molecular absorption lines at high redshift can provide an excellent
probe of cosmological physics such as the cosmic microwave background,
values of the fundamental constants and the chemistry of the early
Universe \citep[e.g.][]{wc96a,wc97,wc01,dwbf98}. However, such studies
are limited to the 4 known high redshift molecular absorption systems,
towards TXS 0218+357 \citep{wc95}, PKS 1413+135 \citep{wc97}, TXS
1504+377 \citep{wc96} and PKS 1830--211 \citep{wc98}. In the search
for new systems, one systematic approach is to target high column
density absorbers with known redshifts. A convenient sample is the
damped Lyman-alpha absorbers, which have neutral hydrogen column
densities $N_{\rm HI}\ga10^{20}$ cm$^{-2}$. In order to select a
sample, we produced a catalogue of all known DLAs
\citep{cwbc01}\footnote{A version of this catalogue is continually
updated on-line and is available from
http://www.phys.unsw.edu.au/$\sim$sjc/dla} and shortlisted those which
are illuminated by radio-loud quasars (i.e. those with a measured
radio flux density $>0.1$ Jy). From this sample of 57 we selected
those which have 12 mm or 3 mm fluxes.

Recently, we completed a search for molecular absorption towards 11
DLAs at $\lambda\leq3$ mm with the SEST 15-m and Onsala 20-m
telescopes which, apart from one tentative detection, only lead to
upper limits for 18 transitions \citep{cwn+02}. With the upgraded ATCA it is
now possible to improve on these previous attempts. In this paper we
present the results of our first search with this
telescope -- observations towards the known southern centimetre-loud
quasars occulted by DLAs in which a commonly detected transition falls
into the 12 mm band.

\section{Observations}\label{subsec:obs}
The observations were performed in June 2002 with the ATCA at
Narrabri, Australia during excellent weather conditions which gave
good phase stability. The telescope has recently been upgraded with
the addition of 12 mm and 3 mm receivers to antennae 2, 3 and 4
\citep[see][]{wm02}, thus permitting the search for redshifted
molecular rotational lines with this instrument. As mentioned above,
the 3 mm band has been used quite extensively in previous searches
towards DLAs \citep{wc94a,wc95,wc96b,cwn+02}, although to date no 12
mm searches have been published. At this wavelength we are able to
take advantage of the lower system temperature ($\approx60$ K) and
better atmospheric stability to observe towards the southern
($\delta\leq30$\dg) quasars of sufficient flux\footnote{Originally PKS
0336--017 was confused with Q 0336--019 (J 0339--017), which was then
used as a calibrator for the source (Table \ref{t1}). The measured 22
GHz flux density of $S_{22}=0.15$ Jy now joins $S_{0.4}=1.31$,
$S_{1.4}=0.60$, $S_{2.7}=0.45$ and $S_{5.0}=0.30$ Jy for the measured
radio flux densities of 0336--017 \citep{cwbc01}.} occulted by DLAs at
redshifts of $\sim3$ (Table \ref{t1})\footnote{HCO$^+$ is the
strongest and most commonly detected molecule in the four known high
redshift molecular absorption systems (see Wiklind \& Combes
references). Note that no CO transitions fell into either of the two
sub-bands at these redshifts.}. In order to minimize the bias
introduced by using such optically selected sources, ideally we would
have selected the most visually faint (and hence dusty) objects. However,
the ATCA still has a relatively narrow tuning range in the 12 mm band.
We therefore selected all of the ``high flux'' DLAs
which can be observed with this restriction (see Table \ref{t1}). Fortunately, all
three are relatively faint (by DLA standards) with 0201+113 and
0537--286 also being red (${\rm B}-{\rm R} = 2.9$ and 0.9, respectively).
\begin{table*}
	\centering
	 \begin{minipage}{162mm} 
\caption{The 22 GHz illuminated DLAs where the redshifted HCO$^+$
$0\rightarrow1$ falls into either the 16.089--18.888 or 20.089--22.488
GHz ATCA bands. $z_{\rm abs}$ is the DLA redshift and $\nu_{\rm obs}$
is the expected frequency of the redshifted HCO$^+$ $0\rightarrow1$
line, given to the number of significant figures available from the
optical data. The red, visible and blue magnitudes of the quasar are
given as well as the measured 22 GHz flux densities
from both the literature (Lit.) and our observations (Obs.). Any
difference between these two values may be due to quasar variability.}
\begin{tabular}{@{}l c c c c r c c l c l c c@{}} 
\hline
Source & \multicolumn{2}{c}{Coordinates (J2000)} &   $z_{\rm abs}$ &
$\nu_{\rm obs}$& \multicolumn{3}{c}{Magnitude} &\multicolumn{2}{c}{$\approx S_{22}$ (Jy)} & Calibrator &\multicolumn{2}{c}{$\approx S_{22}$ (Jy)}\\ & h ~m ~s & d~~~$'$~~~$''$ &  & & B & V & R & Lit.\footnote{The flux density for 0201+113 is from \protect\citet{tuwv01} and for 0202+149, 0336--019 and 0537--286 from the ATCA calibrators site (http://www.narrabri.atnf.csiro.au/calibrators/). The uncertainty of $\approx15$\% in these calibrators is the major contributor to the errors in the measured flux densities.}& Obs.& &Lit.  & Obs.\\ 
\hline 
0201+113 & 02 03 46.7 & 11 34 44 & 3.38639 & 20.331 &  21.7 & 19.5 & 18.8 & 0.55 & $0.59$ & 0202+149&2.00 &
$1.41$\\ 
0336--017 & 03 39 00.9 & -01 33 18 &  3.0619 & 21.958& 19.5 & 18.8 & 19.1 &-- & $0.15$ & 0336--019&3.37 &$2.2$\\ 
0537--286 & 05 39 54.3 & -28 39 56 & 2.976 & 22.43 & 19.8 & 19.0 & 18.9 & 1.66 & $0.58$&
\multicolumn{3}{c}{\it Self calibration}\\ 
\hline
\end{tabular}
\label{t1}
\end{minipage}
\end{table*}


Because of the lack of precision in the optical redshifts (most
significant for 0537--286, Table \ref{t1}), we sacrificed one
polarisation in order to overlap two of the widest available (64 MHz)
bands. This enabled us to cover an uncertainty of $\approx\pm 0.01$ in
redshift while retaining a relatively high spectral resolution of 3.4
\kms. Finally, post observation, baseline 2--4 was discarded because of
phase referencing problems. No other flagging of bad data was
required. For 0201+113 and 0336--017 the bandpass of the calibrator
was removed from the spectra and in the case of 0537--286, which was
self calibrated, we removed a low order polynomial to ``flatten'' the
bandpass.

\section{Results and Discussion}

\begin{figure}
\vspace{5.4 cm} \setlength{\unitlength}{1in} 
\begin{picture}(0,0)
\put(-0.52,2.7){\includegraphics{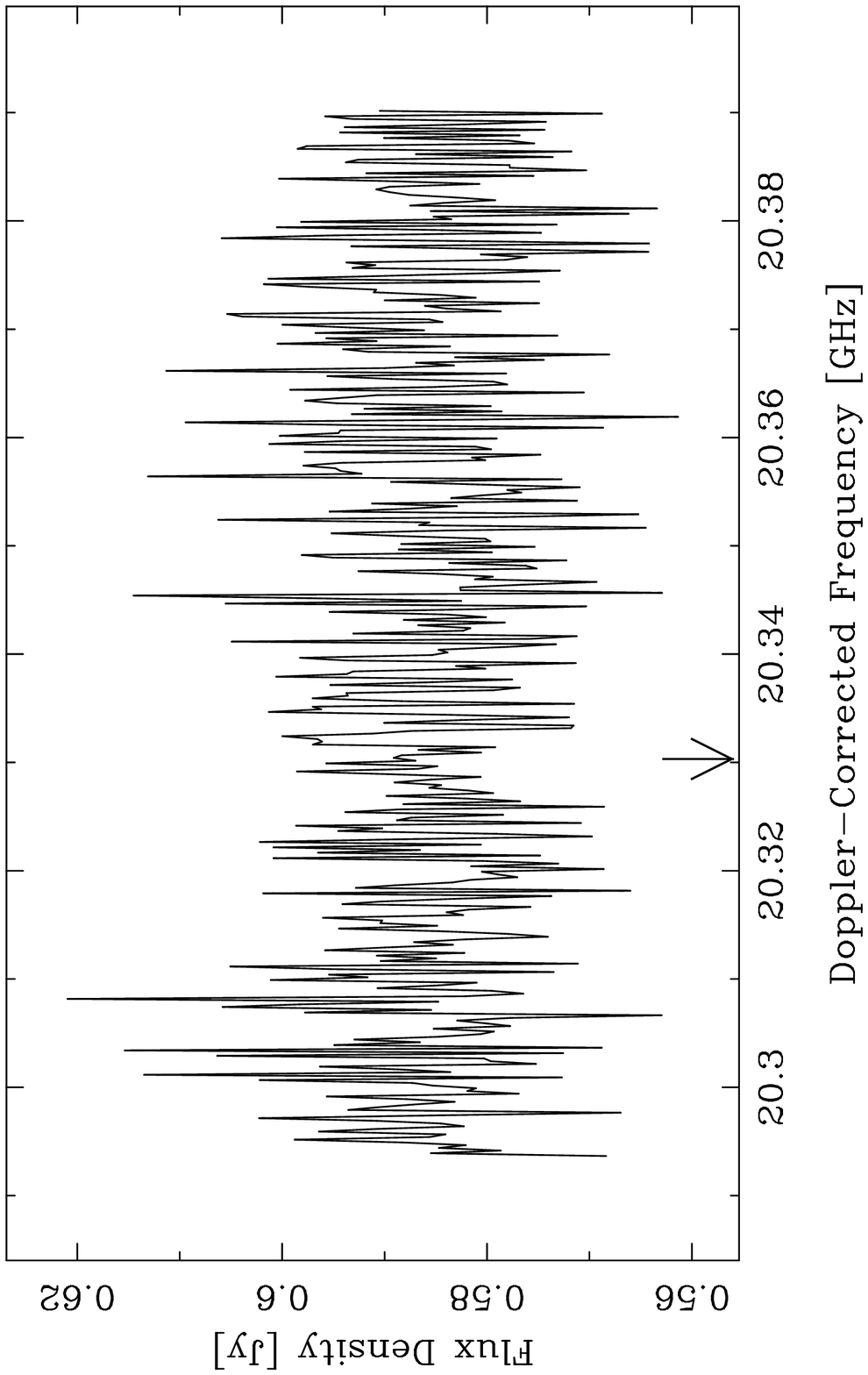}}
\end{picture}
\caption[]{HCO$^+$ $0\rightarrow1$ at $z=3.386\tiny\begin{array}{c}{+0.013}\\{-0.009}\end{array}$\normalsize towards 0201+113. The $1\sigma$ r.m.s. noise is 10 mJy. The overlap between the two spliced bands is 14 MHz. In this and Figs. \ref{f2} and \ref{f3} the arrow shows the expected absorption frequency according to the published DLA redshift (Table \ref{t1}).}
\label{f1}
\end{figure}

\begin{figure}
\vspace{5.4 cm} \setlength{\unitlength}{1in} 
\begin{picture}(0,0)
\put(-0.52,2.7){\includegraphics{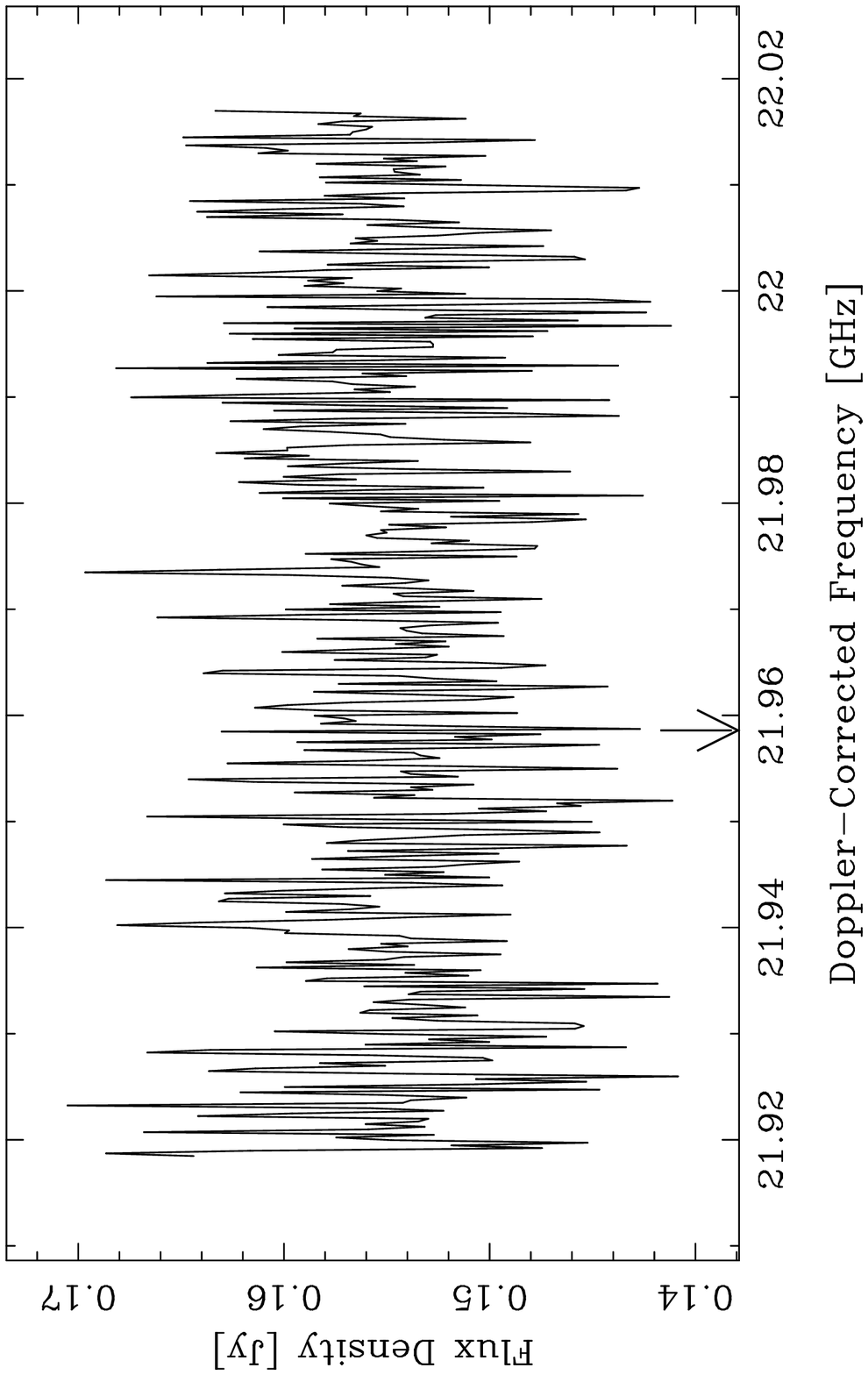}}
\end{picture}
\caption[]{HCO$^+$ $0\rightarrow1$ at $z=3.062\tiny\begin{array}{c}{+0.011}\\{-0.007}\end{array}$\normalsize towards 0336--017. The $1\sigma$ r.m.s. noise is 6 mJy. The overlap between the two spliced bands is 12 MHz.}
\label{f2}
\end{figure}

\begin{figure}
\vspace{5.4 cm} \setlength{\unitlength}{1in} 
\begin{picture}(0,0)
\put(-0.52,2.7){\includegraphics{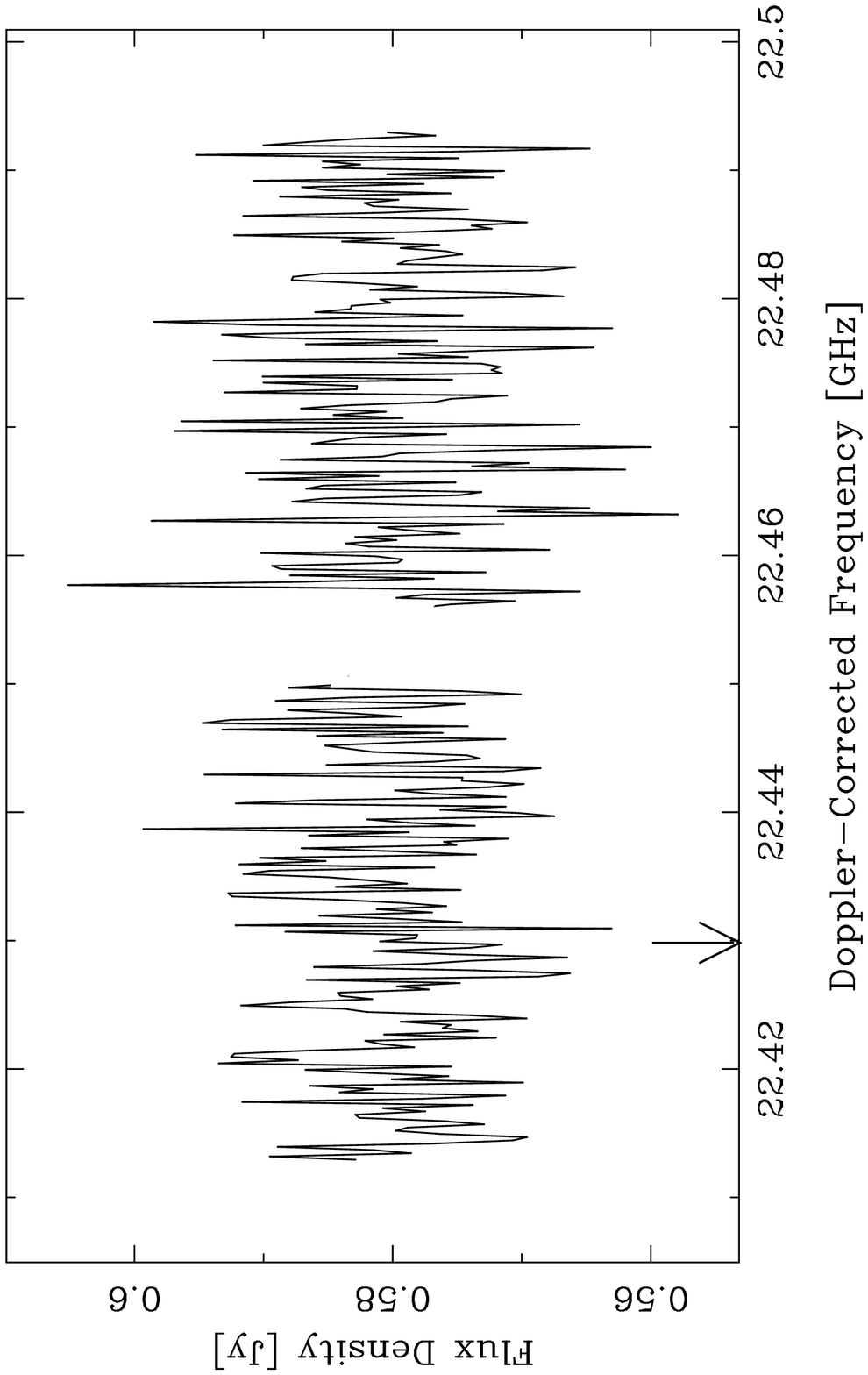}}
\end{picture}
\caption[]{HCO$^+$ $0\rightarrow1$ at $z=2.976\tiny\begin{array}{c}{+0.010}\\{-0.004}\end{array}$\normalsize towards
0537--286. The $1\sigma$ r.m.s. noise is 8 mJy. The gap in the
spectrum occurs since there was no bandpass calibration and the fitted
polynomial is unreliable at these frequencies.}
\label{f3}
\end{figure}

In Figs. \ref{f1} to \ref{f3} we show the time averaged spectra
over both good baselines and note that there are
no HCO$^+$ $0\rightarrow1$ absorption features of $\geq3\sigma$ per
3.4 \kms ~channel in these DLAs.
From the r.m.s. noise levels we derive optical depth limits for a
resolution of 1 \kms ~and column density limits according to $N_{\rm
mm}\propto \int\tau dv$ for an excitation temperature of $\approx10$ K
~\citep[see][]{cwn+02}. These are listed in Table \ref{sum} together
with all previously published results.

\begin{table*}
 \centering
 \begin{minipage}{161mm}
\caption{Summary of published searches for rotational molecular
absorption in DLAs. $\nu_{\rm obs}$ is the approximate observed
frequency (GHz), V is the visual magnitude of the background quasar,
$N_{\rm HI}$ (\scm) is the DLA column density from the Lyman-alpha
line and $\tau_{\rm 21~cm}$ is the optical depth of the redshifted 21
cm \HI ~line (see Curran {\rm et~al.} 2002b). The optical depth of the
relevant millimetre line is calculated from $\tau=-\ln(1-3\sigma_{{\rm
rms}}/S_{{\rm cont}})$, where $\sigma_{{\rm rms}}$ is the r.m.s. noise
level at a given resolution and $S_{{\rm cont}}$ is the continuum flux
density. This is done for a resolution of $\Delta v=1$ \kms
~($\tau_{\rm mm}$), where we quote only the best existing limit. Note
that due to simultaneous flux measurements with the ATCA, unlike many
of the other results given, we minimize errors due to variable fluxes
(see Table \ref{t1}). For all optical depths, $3\sigma$ upper limits
are quoted and ``--'' designates where $3\sigma>S_{{\rm cont}}$, thus
not giving a meaningful value for this limit. Blanks in the $\tau_{\rm
21~cm}$ field signify that there are no published \HI ~absorption data
for these DLAs. The penultimate column gives the best existing limit
of the column density per unit line-width (not to be confused with
$\Delta v$) estimated for the transition [\scm (\kms)$^{-1}$].}
\begin{tabular}{@{}l c c r c c c c c c @{}}
\hline
DLA & $z_{\rm abs}$ & Transition & $\nu_{\rm obs}$ & V & $N_{\rm HI}$ & $\tau_{\rm 21~cm}$ &$\tau_{\rm mm}$ & $N_{\rm mm}/dv$ & Ref.\\
\hline
0201+113 & 3.38639 & HCO$^+$ $0\rightarrow1$ & 20.3 & 19.5 &$2\times10^{21}$ &  $0.04-0.09$ &$<0.1$ &$<9\times10^{11}$ & 7\\
0235+1624 & 0.52400 & CO $0\rightarrow1$& 75.6 &15.5 & $4\times10^{21}$ & $0.05-0.5$ & $<0.06$& $<4\times10^{14}$& 2\\
... &0.52398 & CO $1\rightarrow2$ & 151.3& ... & ... & ... & $<0.09$& $<6\times10^{15}$& 4\\
...	&  ...& HCO$^+$ $3\rightarrow4$ & 234.1 & ... &... &...  & $<0.3$ & $<4\times10^{12}$& 4\\
...& 0.523869& CS $2\rightarrow3$ & 96.4 &... & ...&  ...& $<0.9$& $<1\times10^{13}$& 6\\
0248+430 & 0.3939& CS $2\rightarrow3$ & 105.4 &17.7 &$4\times10^{21}$ &  0.20& --&-- & 6\\
0336--017 & 3.0619 & HCO$^+$ $0\rightarrow1$ &22.0 &18.8 &$2\times10^{21}$ &$<0.005$ & $<0.2$&$<2\times10^{12}$ & 7\\
0458--020 & 2.0397  & HCO$^+$  $2\rightarrow3$  & 88.0 &18.4 & $5\times10^{21}$ &0.30  & $<2$&  $<8\times10^{12}$& 5\\
...	&   2.0399 &... & 88.0 & ...& ... & ... & $<0.4$&  $<1\times10^{12}$& 6\\
...	&  2.0397 & CO $2\rightarrow3$ & 113.8 &... & ...&  ... & --& --& 3\\
...	&   2.0399 & CO $3\rightarrow4$& 151.7 &... & ...& ... & $<1$& $<2\times10^{16}$& 6\\
0528--2505 & 2.1408 & CO $2\rightarrow3$ &110.1 &19.0  &$4\times10^{20}$ & $<0.2$ & \multicolumn{2}{c}{{\it No published 3 mm fluxes}} & 3\\
0537--286 & 2.974 & HCO$^+$ $0\rightarrow1$ & 22.4 &19.0 & $2\times10^{20}$ & & $<0.08$ & $<7\times10^{11}$& 7\\
0738+313 &  0.2212 & CO $0\rightarrow1$ & 94.4 &16.1 &$2\times10^{21}$ & 0.07& --& $<4\times10^{15}$  & 6\\
0827+243 & 0.5247 & CS $2\rightarrow3$ & 96.4 &17.3  &$2\times10^{20}$ &0.007  & $<0.4$& $<1\times10^{13}$& 6\\
...	&  0.52476 & CO $0\rightarrow1$ &75.6 & ... &... & ... & $<0.2$ & $<1\times10^{15}$ & 8\\
08279+5255  & 2.97364& HCO$^+$ $0\rightarrow1$ & 44.9 &15.2  &$1\times10^{20}$ &  &\multicolumn{2}{c}{{\it No published 7 mm fluxes}}& 6\\
...	&  ... & CO $2\rightarrow3$ & 87.0 &... &... & ... &\multicolumn{2}{c}{{\it No published 7 or 3 mm fluxes}} & 6\\
0834--201 & 1.715 & HCO$^+$  $2\rightarrow3$ &98.6 &  18.5& $3\times10^{20}$ & & $<0.4$& $<2\times10^{12}$ & 5\\
...	&  ... &HCO$^+$  $3\rightarrow4$& 131.4 &... & ...&  ...&$<0.6$ &  $<8\times10^{12}$& 5\\
...	&  ... &  CO $3\rightarrow4$& 169.8 &... & ...&  ...&--& -- & 6\\
1017+1055 & 2.380& CS $2\rightarrow3$ & 43.5 &17.2 & $8\times10^{19}$ &  &\multicolumn{2}{c}{{\it No published 7 mm fluxes}} & 6\\
...	&  ... & CO $2\rightarrow3$ & 102.3 &... & ...&  ... &\multicolumn{2}{c}{{\it No published 3 mm fluxes}}& 6\\
1215+333 & 1.9984 & CO $2\rightarrow3$ & 115.3 &18.1 & $1\times10^{21}$ &   &--& --  & 3\\
1229--0207 & 0.3950 & CO $0\rightarrow1$  & 82.6 &16.8   &$1\times10^{21}$ &  & $<1$& $<6\times10^{15}$& 6\\
...	&  ... & CO $1\rightarrow2$& 165.3 &... & ...&  ...&--& --& 6\\
...	& 0.39498  & CO $1\rightarrow2$& 165.3 &... & ...&  ...& --& --& 4\\
1328+307 & 0.69215 & HCO$^+$  $1\rightarrow2$ & 105.4 &17.3  &$2\times10^{21}$ &0.11  &$<0.7$ & $<2\times10^{12}$& 4\\
...	&  ... & CS $2\rightarrow3$ & 86.9 &... &... & .... & $<2$& $<4\times10^{13}$& 6\\
...	&  ... & CO $1\rightarrow2$ & 136.2 &... & ...& ... &$<1$ &$<3\times10^{15}$& 4\\
...	&  ... & CO $2\rightarrow3$ & 204.4 &... & ...& ... & --&--& 4\\
1331+170 & 1.7764 & CO $0\rightarrow1$ & 41.5 &16.7 & $3\times10^{21}$ & 0.02 & -- & -- & 1\\
...& 1.7755 & ...&41.5 &  ... & ...& ... & --&--& 1\\
1451--375 &0.2761 & HCO$^+$ $1\rightarrow2$ & 139.8 & 16.7 &$1\times10^{20}$ & $<0.006$ &$<0.6$ & $<2\times10^{12}$ & 6\\
...	&  ... &CO $0\rightarrow1$  &90.3 & ... & ...& ... &$<0.3$ &$<2\times10^{15}$& 6\\
2136+141 & 2.1346 & HCO$^+$ $2\rightarrow3$&85.4 & 18.9  & $6\times10^{19}$&   &$<0.3$ &$<1\times10^{12}$& 5\\
...	&  ... &  CO $2\rightarrow3$ & 110.3 &... &  ...&   ...& $<0.6$& $<4\times10^{15}$&5\\
...	&  ... &  CO $3\rightarrow4$ &147.1 & ... &  ...&   ...&$<0.8$ & $<2\times10^{16}$& 5\\
...	&  ...&  CO $5\rightarrow6$ & 220.6 &... &  ...&  ... & --& --& 4\\
\hline
\end{tabular}
\label{sum}\\
{References: (1) \citet{tsi+84}; (2) \citet{tnb+87}; (3) \citet{wc94}; (4) \citet{wc95}; (5) \citet{wc96b}; (6) \citet{cwn+02};
(7) This paper; (8) A. Bolatto (private communication)
}
\end{minipage}
\end{table*}

From the table we see that, after 3--5 hours per source, our limits
are among the lowest of all the searches and we achieve the most
sensitive search for HCO$^+$ absorption in a DLA published to date.
However, there are uncertainties in the conversion ratio betweeen
$N_{{\rm HCO}^+}$ and $N_{\rm CO}$ (and thus $N_{\rm H_2}$ which the
CO traces\footnote{Although molecular hydrogen constitutes the bulk of
the molecular gas in interstellar space, it cannot be observed
directly due to its small dipole moment and moment of inertia. Since
CO is the next most abundant molecule after H$_2$, this ``tracer'' is
extremely useful in the study of the bulk ISM and is therefore the
most studied molecule in external galaxies.}). The ratio typical of
Galactic star forming clouds is $N_{\rm CO}\gapp10^{4} N_{{\rm
HCO}^+}$ \citep[e.g.][]{wc95}, whereas that for Galactic absorbers
towards extragalactic continuum sources\footnote{This applies in the regime in
which we are interested, i.e. CO column densities of $N_{\rm
CO}\lapp10^{15}$ \scm ~(Table \ref{sum}). Above this the CO begins
self-shielding and subsequently $N_{\rm CO}>10^{3} N_{{\rm HCO}^+}$.}
is $N_{\rm CO}\sim10^{3} N_{{\rm HCO}^+}$ \citep{ll98}. For high
redshift clouds, $N_{\rm CO}>10^{3} N_{{\rm HCO}^+}$ \citep{wc95} or,
more specifically, $N_{\rm CO}=7500 N_{{\rm HCO}^+}$ for one of the 4
known absorbers (Section 1), the gravitational lens PKS 1830--211
\citep{mcr99}. Therefore our results are of little use in assigning
upper limits to the molecular hydrogen column density, $N_{\rm
H_2}$. In order to bypass this uncertainty, we use the best $N_{\rm
CO}$ limit (from Table \ref{sum}), to estimate an upper limit, though
we are still forced to assign an expected line-width to a non-detected
feature. For three of the four known high redshift absorbers this is
$\approx20$ \kms ~(the FWHM being $\approx7$ \kms ~towards PKS
1413+135\footnote{See
http://www.phys.unsw.edu.au/$\sim$sjc/dla-fig1.ps.gz},
\citealt{wc94,wc95,wc96b,wc98}). Also, as discussed in \citet{cwn+02},
where non-detections are concerned the spectral resolution of the data
have to be taken into account since these determine the r.m.s. noise
and thus the values of $\tau$ (Table \ref{sum}).

Nevertheless, using the best available optical depth limit (CO
$0\rightarrow1$ towards 0235+1624, Table \ref{sum}) gives $N_{\rm
CO}\lapp7\times10^{15}$ \scm ~for a 20 \kms ~line detected at a
resolution of 1 \kms. This becomes $N_{\rm CO}\lapp3\times10^{15}$
\scm ~for the same (barely resolved) line at 10 \kms ~resolution. So
for $N_{{\rm H}_{2}}\sim10^{4}N_{\rm CO}$ \citep[e.g.][]{wc95}, and
assuming that molecular absorption lines have a FWHM of $\lapp20$
\kms, we can assert that $N_{{\rm H}_{2}}\lapp1\% ~N_{\rm HI}$ for the
DLAs deeply searched for rotational molecular absorption\footnote{Note
that the $N_{{\rm H}_{2}}/N_{\rm CO}$ ratio may exceed $10^5$
\citep{ll98} which could increase this value to $\lapp10\%$.}.

For high redshift ($z>1.8$) sources, the ultra-violet H$_2$ lines are
redshifted to optical wavelengths and indeed H$_2$ has been found in
absorption in 7 DLAs at $N_{{\rm H}_{2}}/N_{\rm HI}$ ratios of
$\leq1$\%
\citep{gb97,sp98,lmc+00a,psl00,lddm01,lsp02}. In the low metallicity
environments typical of DLAs ([Zn/H]~$\approx-2$ to $-1$, i.e. $\sim$
0.01 to 0.1 solar, e.g. \citealt{pgw01}), \citet{lis02} argues that
even if the gas is cool, the ambient ionization is sufficient to
suppress the formation of H$_2$. This may offer a potential explanation
for our non-detections.

The under-abundance of heavier elements introduces another
uncertainty: the application of Galactic CO-to-H$_2$ conversion
ratios. \citet{bcf87} and \citet{cbf88} suggest that the
photoionization of carbon would reduce the ratio of $N_{\rm H_2}$ to
$N_{\rm CO}$ to a tenth of Galactic values and, from an upper limit of
$N_{\rm CO}$ from the A--X molecular bands, \citet{sp98} find $N_{\rm
CO}/N_{\rm HI}<10^{-8}$ in DLAs, which is indeed only 10\% of the
local ratio. In addition, at earlier epochs of chemical enrichment we
would expect that there is simply a lack of raw materials to form
significant amounts of tracer molecules (e.g. CO, HCO$^+$, HCN and CS)
even in cold dark clouds where H$_2$ readily forms. A reliable
conversion ratio could be obtained by searching for CO absorption
towards the DLAs in which H$_2$ has been detected. However, the
detection of H$_2$ biases towards optically bright sources and hence
an under-abundance of dust. Furthermore, none of the quasars
illuminating the 7 known H$_2$ absorbing DLAs have appreciable 12 or 3
mm fluxes (Curran et al., in preparation).

In summary, not only do the low metallicities reduce the molecular
hydrogen content in DLAs, but they may also further lower the column
densities of other molecular tracers, making the detection of these
much more difficult than expected.  It is clear that further
statistics are required in order to identify the factors most crucial
(e.g. column density, visual magnitude or metallicity) in determining
the molecular content of DLAs. It would be of interest
to compare the metallicities of the four known molecular absorbers,
especially in the highest redshift case ($z_{\rm abs}=0.886$,
\citealt{wc98}). However, these lie along the lines-of-sight to quasars
which, although bright in the millimetre regime, are too optically dim
(V$>20$ in all cases) to estimate the metal-to-hydrogen ratio.

\section*{Acknowledgments}
We wish to thank the John Templeton Foundation for supporting this
work. Also Tony Wong and Lister Staveley-Smith at the Australia
Telescope National Facility for their support in using the upgraded
ATCA and Alberto Bolatto at the Berkeley Radio Astronomy Laboratory
for his preliminary results. Steve Curran acknowledges receipt of a
UNSW NS Global Fellowship. This research has made use of the NASA/IPAC
Extragalactic Database (NED) which is operated by the Jet Propulsion
Laboratory, California Institute of Technology, under contract with
the National Aeronautics and Space Administration.


\bsp

\label{lastpage}

\end{document}